\documentclass[twocolumn,showpacs,preprintnumbers,amsmath,amssymb]{revtex4}

\usepackage[dvips]{graphicx}
\usepackage{dcolumn}
\usepackage{bm}

\begin{document}


\title{Low frequency noise controls on-off intermittency of bifurcating systems}

\author{S\'ebastien Auma\^{\i}tre, Fran\c{c}ois P\'etr\'elis}

\affiliation{
Laboratoire de Physique Statistique, Ecole Normale Sup\'erieure, 24
rue Lhomond, 75005 Paris, France
}

\author{Kirone Mallick}
\affiliation{SPhT, Orme des Merisiers, CEA-Saclay, 91191 Gif-sur-Yvette, France}%

\date{\today}

\begin{abstract}
A bifurcating system  subject to  multiplicative noise can display on-off intermittency. Using a canonical example, we investigate the extreme sensitivity of the intermittent behavior to the nature of the noise. Through a perturbative expansion and  numerical studies of the probability density function of the unstable mode, we show that intermittency is controlled by the ratio between the departure from onset and the value of the noise spectrum at zero frequency. Reducing the noise spectrum at zero frequency shrinks the intermittency regime drastically. This effect also modifies the distribution of the duration that the system spends in the off phase. Mechanisms and applications to more complex bifurcating systems are discussed.

\end{abstract}


\maketitle

Among the possible behaviors of a chaotic system  is  
intermittent behavior. The system remains for long durations in some
regular state (say a laminar state or off-phase) and  at unpredictable instants
begins to explore other states (say on-phase) before returning to the laminar state.

A simple deterministic model for intermittency was proposed by Pomeau
and Manneville \cite{Pomeau}: a limit cycle is weakly unstable but
from time to time a reinjection mechanism forces the system to
return close to this limit cycle. 
A few years later, a new
type of intermittency was discovered in coupled dynamical systems \cite{japonaisonoff} and also  identified in a system of
reaction-diffusion equations \cite{pikovskyonoff}. 
Both systems can be approximately described by the evolution of a
weakly linearly unstable mode with a noisy control parameter. 
This type of intermittency was given the name ``on-off intermittency''
by Platt, Spiegel and Tresser \cite{spiegel} who pointed out its
genericity when an unstable system  is coupled to a system
that evolves in an unpredictable manner.  
Experimentally, on-off intermittency has been identified in various systems including  electronic devices,  electrohydrodynamic convection in nematics,  gas discharge plasmas and spin-wave instabilities  \cite{exptot}.

It is surprising  that, despite the genericity of the
on-off intermittency mechanism,  this effect has not been  
reported more often. One might expect that any careful experimental
investigation of an instability should reveal on-off intermittency when
the system is close to the onset of instability, and is hence sensitive
to  unavoidable experimental noise in the control parameters. This remark is the main
motivation for the present work. We  show that the amplitude of
the noise is not the relevant control parameter of  on-off
intermittency. 

Through an
analytical study of a simple stochastic system, we  identify the 
parameter that drives the intermittent behavior and compare  our
prediction to numerical simulations. We then test our prediction for  a chaotic rather than a stochastic system. Then we  discuss the sensitivity
of the statistics  of the laminar phase duration to the parameter that controls the on-off intermittency.  Finally, we   present applications of this result to   complex systems.

One of the simplest systems that can exhibit on-off intermittency is 
\begin{equation}
 \dot{X}=(a+\zeta(t)) X-X^3\,,
\label{eqbase}
\end{equation}
where $\zeta$ is a random process with zero mean \cite{japonaisonoff}. In the deterministic
regime (no-noise), the variable $X$ undergoes a pitchfork supercritical
bifurcation for $a=0$. The attractor, $X=0$, is stable for negative $a$
and is
unstable for positive $a$: $X$ tends in the long time limit to one of its two
stable attractors $\pm \sqrt{a}$. 
In the stochastic regime, the noise $\zeta$ acts as a modulation of the forcing
parameter.  
Note that if the 
initial condition verifies $X(t=0)\ge0$ then $X(t)\ge0$ for  all time. Henceforth we 
consider only positive initial conditions for $X$ without any loss of generality.

For stationary Gaussian white noise,  the probability density
function (PDF) of $X$ is derived by solving the associated  Fokker-Planck
equation  \cite{Graham}. We 
define the noise intensity by
$\langle \zeta(t)\zeta(t')\rangle_s= D \delta(t-t')$ where 
$\langle\rangle_s$ is the average over realizations of the noise. Equation (\ref{eqbase}) is then understood in the sense of Stratonovich. 
If $a\le 0$, $X$
tends to zero and the stationary PDF is $P(X)=\delta(X)$. For positive $a$, one  obtains 
\begin{equation}
P(X)=C X^{\frac{2 a}{D}-1}e^{-\frac{X^2}{D}}\,,
\label{pdfgaussian}
\end{equation}
where $C$ is a normalization constant. 
For $0\le 2 a/D < 1$, this PDF
diverges at the origin $X=0$. As pointed out in \cite{japonaisonoff},  this divergence is associated with the intermittent behavior of $X$, as $X$
  remains for long durations arbitrarily close to the unstable fixed
point $X=0$. When $2 a/D$ is large,  intermittency disappears and $X$
fluctuates around its deterministic value, $\sqrt{a}$.

\begin{figure} 
\includegraphics[width=7cm]{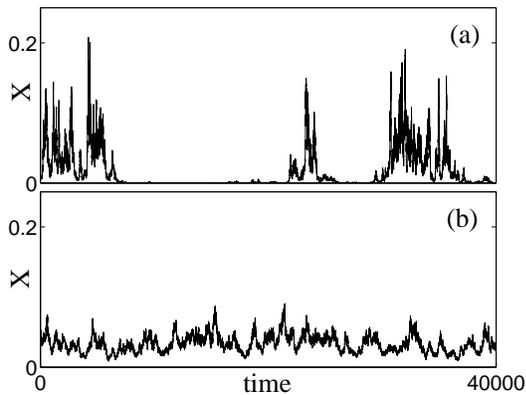}
\caption{ Temporal traces of the dynamical variable $X(t)$, solution of 
 (\ref{eqbase}), with $a=1.25\times 10^{-3}$, $\alpha^2=0.005$ and  the noise defined by (\ref{selfcorrbr}): 
(a)  $\eta=\Omega=0.25$ {\it i.e.}, $a/S=0.3927$; (b)  $\eta=\Omega=2.5$ {\it i.e.}, $a/S=3.9270$.
\label{fig1}}
\end{figure}

However,  for  colored noise, a more complex situation is expected. It is tempting to assume that the noise amplitude $\sqrt{\langle \zeta(t)^2 \rangle_s}$ controls  the on-off regime. This is not the case. 
We plot in fig. \ref{fig1} the solution of equation
(\ref{eqbase}) for two different colored noises with the same value of 
$\langle\zeta^2\rangle_s$. One of the solutions is intermittent but not 
 the other. Hence intermittency is controlled by another parameter of the system.

In order to identify this parameter, 
 we derive  an approximate expression of the stationary PDF of $X$ using the
cumulant expansion introduced by van Kampen 
 and shown to be valid for small values of $\alpha \tau_c$ where $\alpha$
is the noise amplitude and $\tau_c$ its correlation time \cite{vankampen2}. In the case under study,  
two parameters appear in the expansion that are related to the autocorrelation function:
\begin{equation}
S=\int_0^\infty \langle \zeta(0)\zeta(\tau)\rangle_s d\tau \,\,,\,\,
M=\int_0^\infty \langle\zeta(0)\zeta(\tau)\rangle_s e^{-2 a \tau} d\tau\,.
\label{defSM}
\end{equation}
For a Gaussian white-noise, the expansion is exact and leads to the stationary
PDF given by equation (\ref{pdfgaussian}).  
In the following we consider the generic case where $S$ and $M-S$ are non-zero.  The stationary PDF is
\begin{equation} 
P(X)=C X^{\frac{a}{S}-1} |1+\frac{(M-S)X^2}{S\, a}|^{-(1+\frac{a\,M}{2\,S\,(M-S)})}\,,
\label{pdfcolored}
\end{equation}
where $C$ is a normalization constant. 
The behavior of the PDF for $X$ close to zero is proportional to
$|X|^{\frac{a}{S}-1}$. 
It diverges  for $X=0$ so that  
on-off intermittency occurs if
\begin{equation}
0<\frac{a}{S}<1\,.
\label{eqcriterium}
\end{equation}
This is consistent with fig.\ref{fig1} since  $a/S=0.3927$ for the intermittent signal
and $a/S=3.927$ for the other one. The Wiener-Khintchin theorem
states that  the integral of the correlation function, $2 S$, is equal to the spectrum of the noise at zero frequency.   
Thus, another interpretation of the
criterion (\ref{eqcriterium}) is that on-off intermittency is present
when the departure from the deterministic onset is smaller than half  the 
value of the noise spectrum at zero frequency.    
This interpretation also holds  when the noise is white and Gaussian since $S=D/2$. For white-noise, the spectrum has the same value, $D$, for all the frequencies. Consequently, the analytical result (\ref{pdfgaussian}) does not identify which part of the spectrum  controls the intermittency.

\begin{figure}[!htb]
\centerline{\includegraphics[width=7cm]{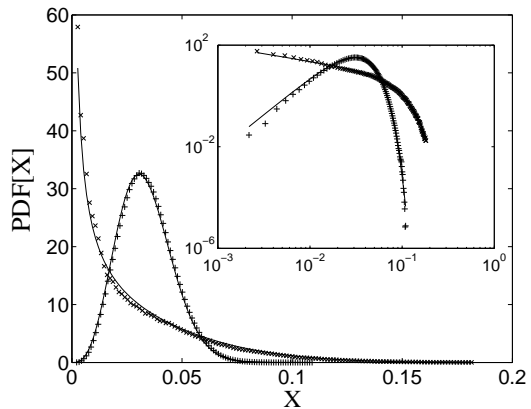}}
\caption{PDF of the solutions of  (\ref{eqbase}) with  the 
noise (\ref{selfcorrbr}). The symbols ($\times$) and ($+$) correspond, respectively,  to the parameters used  in figures 1a and 1b.   
The full lines are the corresponding
predictions given by (\ref{pdfcolored}). Same figure in inset using log-log scale.
\label{fig2}}
\end{figure}

\begin{figure}[!htb]
\centerline{\includegraphics[width=7cm]{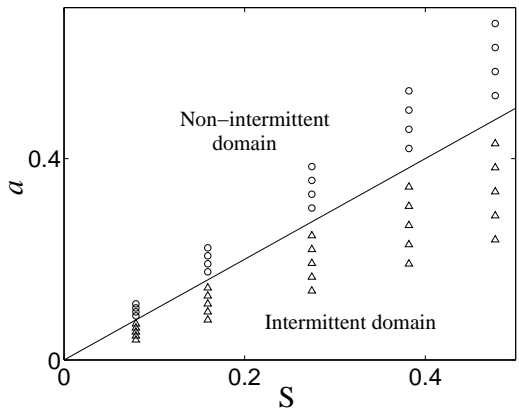}}
\caption{Behavior of the solution of  (\ref{eqbase}) for  noises defined in (\ref{selfcorrbr}) and associated with various values of $S$ and $M$.  ($\triangle$) : intermittent behavior, ($\circ$) :  non-intermittent behavior. The full line is the transition curve predicted by (\ref{eqcriterium}).
\label{fig2b}}
\end{figure}

In order to check the validity of this result,
we solve equation (\ref{eqbase}) numerically, using a stationary Gaussian correlated noise $\zeta(t)$  with autocorrelation function 
\cite{sawford} :
\begin{eqnarray}
\langle \zeta(t) \zeta(t+\tau) \rangle_s=\alpha^2 &
\left(\cos(2\pi \Omega \tau)
+ \frac{\eta}{\Omega} \sin(2\pi \Omega |\tau|)\right)\nonumber\\ &  \exp{(-2\pi \eta |\tau|)}\,.
\label{selfcorrbr}
\end{eqnarray}
The noise variance is $\alpha^2$ and its correlation time $\tau_c=(2\pi \eta)^{-1}$.
This provides   
$S=\alpha^2  \eta/\left[\pi(\eta^2+\Omega^2)\right]$ and 
$M=\alpha^2 (\eta+a/(2 \pi))/\left[\pi\left((\eta+a/\pi)^2
+\Omega^2\right)\right]$. Therefore by changing $\eta$ and
$\Omega$ we can tune   $a/S$ and $\alpha\, \tau_c$ independently. Gaussian white noise is recovered in the limit $\eta \rightarrow {\infty}$ 
with  $\alpha^2/\eta=D$. 
Figure \ref{fig1} presents time series of $X$ 
and  fig. \ref{fig2} the corresponding PDFs. 
We also compare the predicted criterion for appearance of intermittency (\ref{eqcriterium}) with the numerical results. To wit, we draw in figure \ref{fig2b} a phase diagram  in the  $(S,a)$-plane using noises with different $S$ and $M$.  We calculate $X_{mp}$ the most probable value of $X$. For $a>0$,  the solution $X=0$ is unstable. The system is intermittent if $X_{mp}=0$, and  non-intermittent if $X_{mp}\neq 0$. 
In these figures and for all tested parameter values for which $\alpha \tau_c$ is small, there is  a very good agreement between  the numerical results and the predictions (\ref{pdfcolored},\ref{eqcriterium}).

Up to now, we have only dealt with fluctuating parameters that are
random processes. It is tempting to test the prediction
(\ref{pdfcolored},\ref{eqcriterium}) with a deterministic but chaotic fluctuating parameter.
Thus, we study equation (\ref{eqbase}) when $\zeta$ is obtained from the chaotic solution of the Lorenz system \cite{lorenz}, {\it i.e.},  we 
solve
\begin{equation}
\dot{U}=-\sigma(U-Y)\,,\dot{Y}=r U-Y-U Z\,,\dot{Z}=U Y-b Z\,,
\label{eqlorenz}
\end{equation}
and define $\zeta$ by
\begin{equation}
\zeta=\alpha\frac{\mu\dot{U_n}+(1-\mu) U_n }{\langle (\mu \dot{U_n}+(1-\mu) U_n)^2\rangle^{1/2}}\,,
\label{defzeta}
\end{equation}
$U_n=\frac{U-\langle U \rangle}{\sqrt{\langle(U-\langle U \rangle)^2\rangle}}$ and
$\dot{U_n}=\frac{\dot{U}-\langle\dot{U}\rangle}{\sqrt{\langle(\dot{U}-\langle\dot{U}\rangle)^2\rangle}}$. Averages are now understood as long time averages  
 and $\sqrt{<\zeta^2>}=\alpha$. The parameter $\mu$ is tuned  between zero and
one to change the amplitude of the noise spectrum at
zero frequency. Since $\dot{U}$ is the derivative of $U$, the value of its spectrum at low frequencies is smaller than  that of $U$. Increasing $\mu$ increases the
weight of $\dot{U}$ and thus decreases the noise spectrum at low 
frequencies (accordingly the value of $S$).

\begin{figure}[!htb]
\includegraphics[width=8cm]{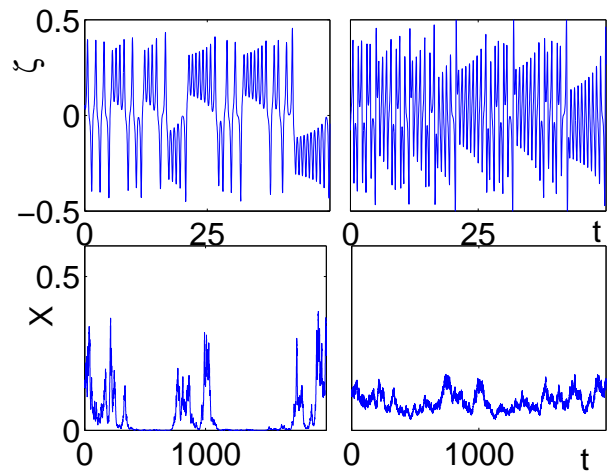}
\caption{Bottom panels: solutions of (\ref{eqbase}) with $\zeta$ obtained from the Lorenz system  (\ref{defzeta}). The chaotic functions $\zeta$ are displayed in the top panels. The parameters are 
$a=0.01$,  $\sqrt{<\zeta^2>}=0.2$ and $\mu=0$ (left panels),  $\mu=0.8$ (right panels). 
Note the difference in the time scales $t$ between the top and bottom panels.
\label{fig3}}
\end{figure}

\begin{figure}[!htb]
\centerline{\includegraphics[width=7cm]{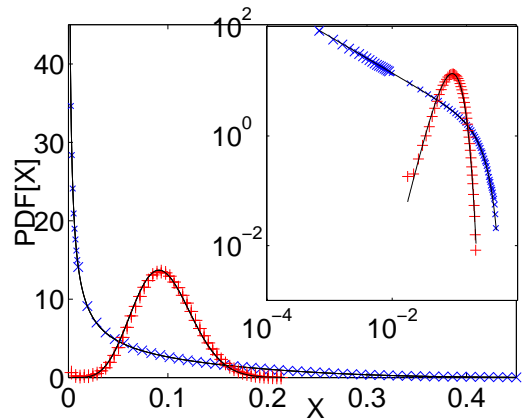}}
\caption{PDF of the solutions of  (\ref{eqbase}) with $\zeta$  obtained from the Lorenz system  (\ref{defzeta}). The parameters are the same as in figure \ref{fig3} with ($\times$) $\mu=0$,  ($+$) $\mu=0.8$.  The inset is the same figure with log-log scales. The full lines are the corresponding predictions  (\ref{pdfcolored}). 
\label{fig4}}
\end{figure}

The equations (\ref{eqbase}, \ref{eqlorenz}) are then solved numerically for $r=25$, $\sigma=10$
and $b=8/3$. The solution of equation (\ref{eqlorenz}) is  chaotic
and we plot  examples of time series of $\zeta$ and
$X$ in fig. \ref{fig3}. As expected  on-off intermittency disappears when $\mu$ increases, {\it i.e.}, $S$
decreases. 
For the
intermittent signal we have $a/S\simeq0.332$, but $a/S\simeq 5.64$ for the non-intermittent one. The numerical estimates of the PDFs of $X$  are compared to the fit
given by (\ref{pdfcolored}) 
(in fig. \ref{fig4}). Again, for small values of the noise amplitude, the
agreement between the prediction and the numerical results is very good.

Interesting results can also be obtained for the duration of the laminar phases $\tau$.  We define a laminar phase as follows: $X(t_0)=\epsilon$, $X(t)< \epsilon$ for  $t_0 < t < t_0+\tau$  and $X(t_0+\tau)=\epsilon$, $\epsilon$ being an arbitrary threshold below which the system is considered to be in the laminar state.   
 Close to the onset of on-off
intermittency, the probability $P(\tau)$ of the duration  $\tau$ of the laminar  phase satisfies $P(\tau)\propto \tau^{-3/2}$ \cite{heagy}. 
 For large values of $\tau$, a cut-off in the power law appears at finite departure from onset  \cite{cenys97}. 
We have checked numerically that for a colored noise, the PDF is indeed proportional to $\tau^{-3/2}$ with a cut-off for high $\tau$. The position of the cut-off increases when $S$ increases ($a$ being constant). This is consistent again with our interpretation of the role of the noise spectrum at zero frequency: the smaller $a/S$ is, the longer the system remains in the laminar state and the more intermittent the signal appears.

Our interpretation of the phenomenon is as follows. On-off intermittency occurs because of a competition between the noise and a systematic drift driven by the  distance from onset. More precisely, as pointed out in \cite{japonaisonoff}, when $X$ is close to the unstable manifold $X=0$, the evolution of $Y=\log{X}$ is given by $\dot{Y}=a+\zeta(t)$. For positive $a$, $\dot{Y}$ is positive on  average  but events in which $Y$ remains smaller than its initial value are possible provided $I=\int_0^T \zeta(t) dt /T$ remains smaller than $-a$ for a long duration. In the  long time limit, the main contribution to the integral $I$ is due to the zero frequency component of the noise. If this component is reduced, then occurrences of the inequality $I\le-a$ are less probable and intermittency tends to disappear. Note that our analytical calculations are based on perturbative expansions and are valid for small values of the product of the noise amplitude with its correlation time. However, even for a  finite amplitude of  noise, we have verified that when the low frequencies  are filtered out, intermittency disappears.

We expect that the role of the zero frequency component of the noise is generic and  also pertinent for systems more complex than the one presented here.  For instance, L\"ucke and Schanck studied a system similar to eq.(\ref{eqbase}) with inertia taken into account. Through a perturbative expansion close to the deterministic solution, they calculated a noise-induced postponement of the onset of instability and a modification in the amplitude of the unstable mode. As they pointed out later, their expansion is not correct when the noise spectrum does not vanish at zero frequency  \cite{Lucke85}. Recent calculations for the same system have shown that for a small departure from  onset and for an Ornstein-Uhlenbeck or a white noise, the PDF of the unstable mode diverges close to zero  \cite{Kirone}. In the light of our study, both the  divergence of the PDF  and the failure of the  perturbative expansion are related to the same physical effect: on-off intermittency when the noise spectrum at zero frequency is non-zero.

Our analysis explains why many experimental investigations on the effect of a multiplicative noise on an instability do not display  on-off intermittency. If the noise is high-pass filtered, as often required for  experimental reasons, then the regime of intermittent behavior disappears. This is the case for instance in \cite{Francois2}: a ferrofluidic layer 
undergoes the Rosensweig instability and peaks appear at the surface. The layer is then subject to a multiplicative noise through random vertical shaking. Close to the deterministic onset, the unstable mode submitted to a colored noise does not display intermittency.

In dynamo theory, the magnetic field is forced by the flow of an electrically conducting fluid. The velocity of the flow appears as a multiplicative term in the equation for the magnetic field. If the flow topology is complex enough and the velocity is large, a magnetic field is generated by dynamo instability. The flow is in general turbulent at dynamo onset so that the velocity fluctuates. We infer that the intermittent behaviors as seen numerically by Sweet {\it et al.} \cite{ott}  are related to the presence of very low frequencies in the spectrum of the velocity field. The same features also occur in simple models of dynamos subject to  white noise \cite{Nico}. On the contrary, experimental realizations of dynamos driven by constrained flows did not display intermittency \cite{expdyn}. These flows, though turbulent, are probably too constrained to display  velocity fluctuations with large enough amplitude of the spectrum at low frequencies.

This work could be generalized to study a parametric instability with a  time-dependent forcing. For an harmonic forcing subject to  frequency or  amplitude noise, intermittent behaviors have been reported \cite{bruitparam}. In these cases, the relevent component of the noise that controls on-off intermittency still remains to be identified.

This work has greatly benefited from discussions with N. Leprovost, N. Mordant, J. Farago, S. G. Llewellyn Smith, S. Fauve and  P. Marcq.

\end{document}